\documentclass[reprint,amsmath,amssymb,aps,prb,groupedaddress]{revtex4-1}
\usepackage{graphicx}
\usepackage{dcolumn}
\usepackage{bm}

\begin{document}
\newcommand{\revtex}{REV\TeX\ }
\newcommand{\classoption}[1]{\texttt{#1}}
\newcommand{\macro}[1]{\texttt{\textbackslash#1}}
\newcommand{\m}[1]{\macro{#1}}
\newcommand{\env}[1]{\texttt{#1}}
\setlength{\textheight}{9.5in}
\bibliographystyle{apsrev4-1}

\title{Creating quantum spin chains through edge reconstruction in pure graphene armchair nanoribbons towards ballistic spin transport}

\author{Ning Wu}
\affiliation{Beijing National Laboratory for Condensed Matter Physics, Institute of Physics, Chinese Academy of Sciences, Beijing 100190, China}
\affiliation{School of Physical Sciences, University of Chinese Academy of Sciences, Beijing 100190, China}
\author{Bang-Gui Liu}\email{bgliu@iphy.ac.cn}
\affiliation{Beijing National Laboratory for Condensed Matter Physics, Institute of Physics, Chinese Academy of Sciences, Beijing 100190, China}
\affiliation{School of Physical Sciences, University of Chinese Academy of Sciences, Beijing 100190, China}

\date{\today}

\begin{abstract}
It is well-known that ferromagnetism can be realized along the zigzag graphene nanoribbon edges, but the armchair graphene nanoribbon edges (AGNEs) are nonmagnetic. Here, we achieve Heisenberg antiferromagnetic spin chains through edge reconstruction along the AGNEs. The reconstructed edge consists of pentagonal carbon rings or a hybrid of pentagonal and hexagonal carbon rings. The resultant nanoribbons are narrow-gap semiconductors and the band edge states are either spin-degenerate edge states or nonmagnetic bulk states. The spin is located on the outermost carbon of the pentagonal ring, and the inter-spin exchange is the nearest-neighbor antiferromagnetic interaction. For finite chain lengthes or nonzero magnetization, there are nonzero spin Drude weights  and thus ballistic quantum spin transport can be achieved along the reconstructed edges, These could be used for quantum spin information transfer and spintronic applications.
\end{abstract}
\maketitle

\section{Introduction}

Graphene nanoribbons (GNRs) have been extensively studied for their exciting properties and potential nanoscale applications\cite{r0,r1,r2,r3,r4}. Their extraordinary electronic and magnetic properties are strongly influenced by their structural boundaries and their widthes and edge geometry structures\cite{r5,r6,r7,r7a,r8}. The most important  pristine edges for GNRs are the zigzag graphene nanoribbons (ZGNRs) and the armchair graphene nanoribbons (AGNRs)\cite{r0}. It has been established that magnetic edges can be achieved in the pristine ZGNRs\cite{r0,r7,zmag1,zmag2}, but no magnetism is found in the pristine AGNRs\cite{r0,r7a}. Further, magnetism can be produced in graphene and GNRs by hydrogen edge modification and doping\cite{r0,gmag1,magH1}, by introducing hetero-atoms\cite{r10,r11} and topological defects\cite{r12,r13}.
On the other hand, although great theoretical and experimental efforts have been devoted to elucidating the real edge structures of GNRs, the issue of whether hydrogen or other functional groups are attached to the edges has not been solved experimentally. The observation of non-functionalized graphene edges in vacuum\cite{r14} provides a strong evidence that the edges of graphene are not always hydrogenated\cite{r15}. Furthermore, while ideal armchair and zigzag edges are the most frequently studied edge types, other specific self-passivating edge reconstructions have been experimentally observed\cite{r16,r17,r18,r19} and have been provided by the global search\cite{r20}. Experimentally, applying the bottom-up approach for the fabrication of GNRs allows rationally designing at the atomic level of both GNRs width and edges state in the past decade\cite{r21,r22,electronics}. Therefore, it is of great interest to explore magnetic properties in AGNRs at the atomic level without external atoms.

Here, we show through first-principles investigation that quantum spin chains can be created in pristine AGNRs by edge structural reconstruction (with carbon only). Such edge reconstruction conserves the non-metallic electronic structures. The reconstructed edges consist of carbon pentagons or a hybrid of hexagons and pentagons, and host quantum Heisenberg antiferromagnetic spin-$\frac12$ chains. These quantum spin chains, with finite lengthes or nonzero magnetization, can be used to transfer spin information due to their super-diffusive and ballistic spin transport\cite{xxz1,xxz2,xxz3,xxz4}. More detailed result will be presented in the following.

\section{Computational method}

Our first-principles calculation is performed using the projector-augmented wave method within the density-functional theory\cite{r23,r24} as implemented in the Vienna Ab initio Simulation Package (VASP)\cite{r25,r26}. The cut-off energy is set to 33 Ry. To describe the exchange-correlation energy, we used the general gradient approximation (GGA) with the Perdew-Burke-Ernkzerhof for solids (PBEsol) parametrization\cite{r27}.
For the computational models, $x$ is the period direction of the ribbon.
To eliminate interactions between the model, the vacuum thicknesses along the $y$ and $z$ direction are both set as 20 {\AA}. In the optimization process, all the C atoms are allowed to relax along all directions, until the force on each atom is less than 0.001 eV/{\AA} and the absolute total energy difference between two successive loops is smaller than 10$^{-8}$ eV. For Brillouin-zone integration, we used a $15 \times 1 \times 1$ grid in the Monkhorst-Pack special $k$-points scheme centered at $\Gamma$ for the geometry optimization, and utilized $25 \times 1 \times 1$ k-point sampling for the self-consistent potential and total energy calculations.
To calculate the electronic band structures of the nanoribbons, 30 k-points were used along the $\Gamma-M$ direction.

\section{Result and discussion}

\begin{figure}[ht]
\centering
\includegraphics[width=0.4\textwidth]{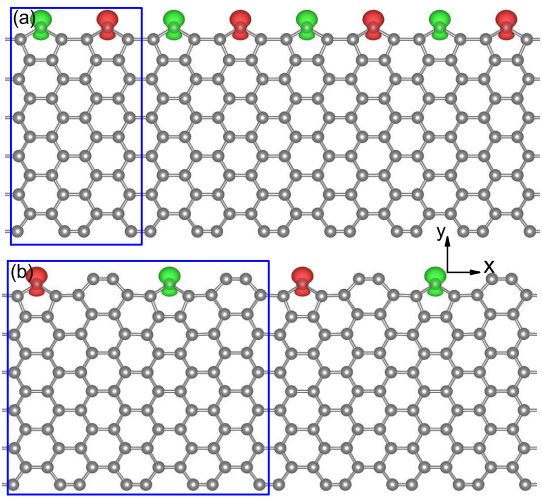}
\caption{\label{Fig1}
The optimized geometrical structures and spin configuration of the pentagonal nanoribbon $AP_{w}$ (a) and the hybrid nanoribbon $AH_{w}$ (b). The blue rectangle defines the unit cells. The magnetization density is described with the green (spin up) and red (spin down) parts.}
\end{figure}

We begin with the pure graphene armchair nanoribbons (ANRs), where both of the edges are terminated with carbon hexagons. We have explored various edge reconstructions to seek carbon spin magnetism from ANRs by constructing various edges and fully optimizing them, and then studying their electronic structures and magnetism. Fortunately, we have found that carbon spin magnetism can be achieved by replacing the carbon hexagons with carbon pentagons at one of the edges, leaving the other edge unchanged, as shown in Figure 1(a). This pentagonal edge can be constructed by putting a single carbon at the bridge point between each pair of carbon hexagons. We use $AP_w$ to denote such an carbon nanoribbon, with one armchair (A) edge and one pentagonal (P) edge, where the subscript $w$ is the width of the nanoribbon. If putting instead a single carbon at every other bridge point, we can obtain a hybrid (H) edge consisting of carbon hexagons and pentagons, still leaving the other edge unchanged, as shown in Figure 1(b). This carbon nanoribbon is denoted with $AH_w$. We use $x$ to denote the direction along the edge, $y$ the perpendicular direction on the nanoribbon plane, and $z$ the plane perpendicular to the $xy$ plane.

In the $x$ direction, the C-C bond length before geometrical optimization in both of $AP_w$ and $AH_w$ is $a_{0}=1.42$ {\AA}, as is extracted from experimental lattice constant for bulk graphene. The corresponding unit cell length $L_x$ is naturally 8.52 {\AA} for $AP_w$ and 17.04 {\AA} for $AH_w$, and they are kept fixed during relaxation to be consistent with the ideal geometry of the graphene sheet in the large $w$ limit. The optimized structural parameters of bond lengthes ($a$, $b$, $c$, $d$, $e$ and $f$) and bond angles ($\gamma$ and $\theta$) (See Figure S1 in the Supporting Information) for $AP_{11}$ and $AH_{11}$ are summarized in Table 1. Compared with the bond length of $a_{0} =$ 1.42 {\AA} in the pristine graphene ANR, the bond lengths of $a_{2}$, $b_{2}$ ($b=2a_0$), $c_{2}$ for $AP_{11}$ and $AH_{11}$ are similar to the bulk bond length values. It is seen that the geometric relaxation near the pentagon edge is quite different from that in the pristine ANR edge. Obviously, structural relaxation has different effects between the pentagon edge and the pristine ANR edge. The bond length $a_1$ for $AP_{11}$ and $AH_{11}$, governing the geometry structure nearby the pentagon rings, is enlarged by 6.90\% and 2.47\% with respect to the C-C bond length of $a_{2}$ values, respectively. In contrast, the bond length $a_3$ is shrunk by 2.11\% and 1.63\% with respect to the C-C bond length $a_{2}$ for $AP_{11})$ and $AH_{11}$, respectively. The bond length $b_1$  between two adjacent carbon atoms near the pentagon edge are compressed, as indicated by $b_{1} < b_{2}$, while the bond length $b_3$ near the pristine edge is enlarged, $b_{2}<b_{3}$. Moreover, it is confirmed that bond lengths and bond angles along the pentagon edge are more significantly changed by the edge reconstruction in $AP_{11}$ than those in $AH_{11}$. The detailed comparison of the C-C bond lengthes and angles between the two edges can help characterize the edge interactions.

Now we determine the magnetic ground state of $AP_{w}$ and $AH_{w}$. The magnetic moment comes from the $p_z$ electron of the outermost carbon of the pentagonal ring along the reconstructed edge, as shown in Figure 1. Total energies ($E_t$) for $AP_{6}$ and $AH_{6}$ in nonmagnetic (NM), ferromagnetic (FM)($\uparrow\uparrow$) and AFM($\uparrow\downarrow$) have been calculated, and the $E_t$ values satisfy $E_t({\rm NM}) > E_t({\rm FM}) > E_t({\rm AFM})$ for the two configurations. It means that AFM state is the most stable configurations for both $AP_{6}$ and $AH_{6}$ cases. In detail, the $E_t$ of the AFM state is 914.24 and 51.84 meV lower than that of the NM and FM states for $AP_{6}$, respectively, and it is 913.75 and 67.37 meV lower than that of the NM and FM states for $AH_{6}$. The total energy relationship holds for other $w$ values (7, 8, 9, 10, 11, 12, 13, and 14), too. Considering that non-magnetic configuration is much higher in total energy than the magnetic ones, we shall show the energy difference between the AFM and FM states in the following.

\begin{figure}[ht]
\centering
\includegraphics[width=0.45\textwidth]{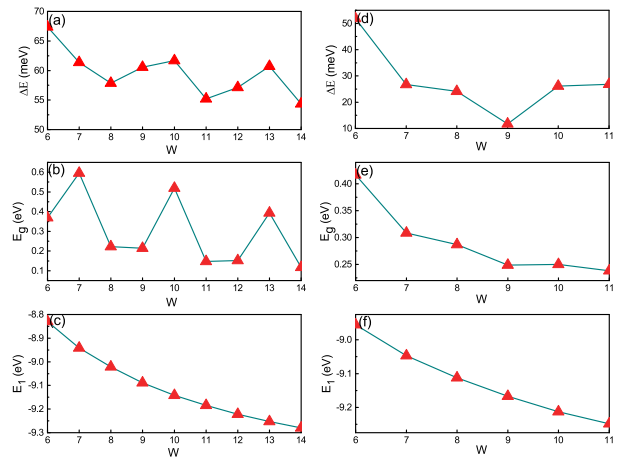}
\caption{\label{Fig2}
The total energy difference ($\Delta E = E_{\rm FM} - E_{\rm AFM}$), band gap ($E_g$), and average ground-state energy ($E_1$,  per carbon atom) of $AP_{w}$ (a, b, c), and those of $AH_{w}$ (d, e, f).  }
\end{figure}

It is very interesting to quantitatively study the effect of ribbon width $w$ on electronic and magnetic properties. By calculating the total energies difference ($\Delta E$) between the AFM and FM states, $\Delta E = E_{\rm FM} - E_{\rm AFM}$, we can demonstrate that the AFM state is the most stable magnetic structure for all the $AP_{w}$ and $AH_w$ cases. The energy gap $E_g$ is also important because we obtain semiconductive energy band structure in all the cases. The structural stability can be characterized by calculating average total energy per carbon atom, $E_1$. For $AP_{w}$, we present in Figure 2 $\Delta E$, $E_g$, and  $E_1$ for $w=6$, 7, 8, 9, 10, 11, 12, 13, and 14. It is interesting that there appears an oscillation with $w$ and the oscillation period for $\Delta E$ appears to be 3 (from $w=7$) and the valley values are located at $w$ = 8, 11, and 14 ($3p+2$ with $p$ an integer), as shown in Figure 2(a). Overall, $\Delta E$ decreases as $w$ increases. Furthermore, the spin exchange coupling can be characterized by one-dimensional Heisenberg antiferromagnetic model,
\begin{equation}
H = J_1\sum \vec{S}_{i}\cdot \vec{S}_{j},
\end{equation}
where $\vec{S}_{i}$ is the spin operator of spin $S=\frac12$ at site $i$, the sum is over $\langle ij\rangle$ pair, and $J_1$ ($=\Delta E$) is the exchange constant. For the $AP_{w}$ with $w= 9$, 10 and 11, the calculated $J_1$ are 60.6, 61.7, and 55.2 meV, respectively. It is implied that the spin AFM order appears along the pure carbon pentagon edge, allowing realization of one-dimensional AFM spin chains with tunable exchange interaction.

For $AP_w$, we present the semiconductor energy gaps ($E_g$) for different width $w$ in Figure 2(b). It is interesting that it is an oscillatory function with the $w$ period of 3, with the top values located at $w = 3p+1$ ($w=7$, 10, and 13), where $p$ is an integer. This width tendency of $E_g$ is similar to that in semiconductor states of H-terminated pristine ANRs from first-principles calculations\cite{r7a,r28,r29}, but distinctive from the tight-binding computations that H-terminated pristine ANR are metallic for $w = 3p+2$, and semiconducting for the others. This is because the tight-binding calculation neglected the effects of the atom relaxation of the ribbon structures\cite{r7} and some technical issues substantially complicated the interpretation of experimental result\cite{r1}. The tunable $E_g$ of $AP_w$ has advantage compared with the traditional semiconductor with fixed $E_g$. In addition, the averaged energy ($E_1$) of the AFM state of the $AP_w$ per carbon atom decreases with the width $w$, as shown in Figure 2(c), indicating the enhancement of stability as the width increased. Here, $E_1$ is defined by $E_{1}=\frac{E_{t}}{N}$, where $N$ is the total number of carbon atoms in the $AP_{w}$ cases.

As for $AH_{w}$ nanoribbons, we have studied the electronic and magnetic properties and present $\Delta E$ in Figure 2(d), $E_g$  in Figure 2(e), and $E_1$ in Figures 2(f) for $w=6$, 7, 8, 9, 10, and  11. Compared to $AP_w$, the width oscillation is still visible  in the $\Delta E$ curve, but no period can be defined for $AH_{w}$. It is clear that $E_g$ also decays with $w$. The averaged energy $E_1$ for $AH_{w}$  also monotonically decrease with the ribbon width $w$ increasing, which is almost the same as in $E_1$ for $AP_{w}$. Because of the total energy relationship $E_t({\rm FM}) > E_t({\rm AFM})$, $\Delta E$ is positive, the most stable configurations of $AH_{w}$ are also in the AFM state. The spin properties can be described by Hamiltonian (1).
For the $AH_{w}$ with $w=9$, 10,  and  11, the calculated $J_1$ per unit cell are 11.8, 26.2, and 26.8 meV, respectively. The spin density for $AH_{11}$ is mainly concentrated at the pentagonal C atoms along the hybrid edge, as shown  in Figure 1(b).

\begin{figure}[ht]
\centering
\includegraphics[width=0.45\textwidth]{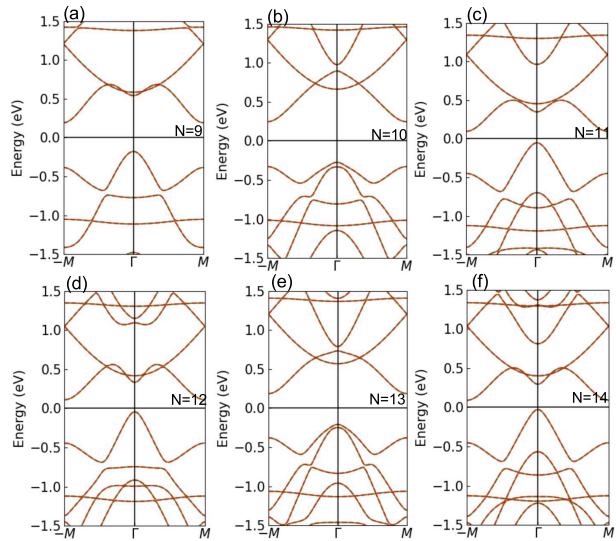}
\caption{\label{Fig3}
The spin-polarized band structures of $AP_{w}$ in the ground-state phase for different ribbon widths:
$w$ = 9 (a), $w$= 10 (b), $w$=11 (c), $w$=12 (d), $w$=13 (e), and $w$ = 14 (f).
The spin-up and spin-down bands are completely degenerate.}
\end{figure}

\begin{figure}[ht]
\centering
\includegraphics[width=0.45\textwidth]{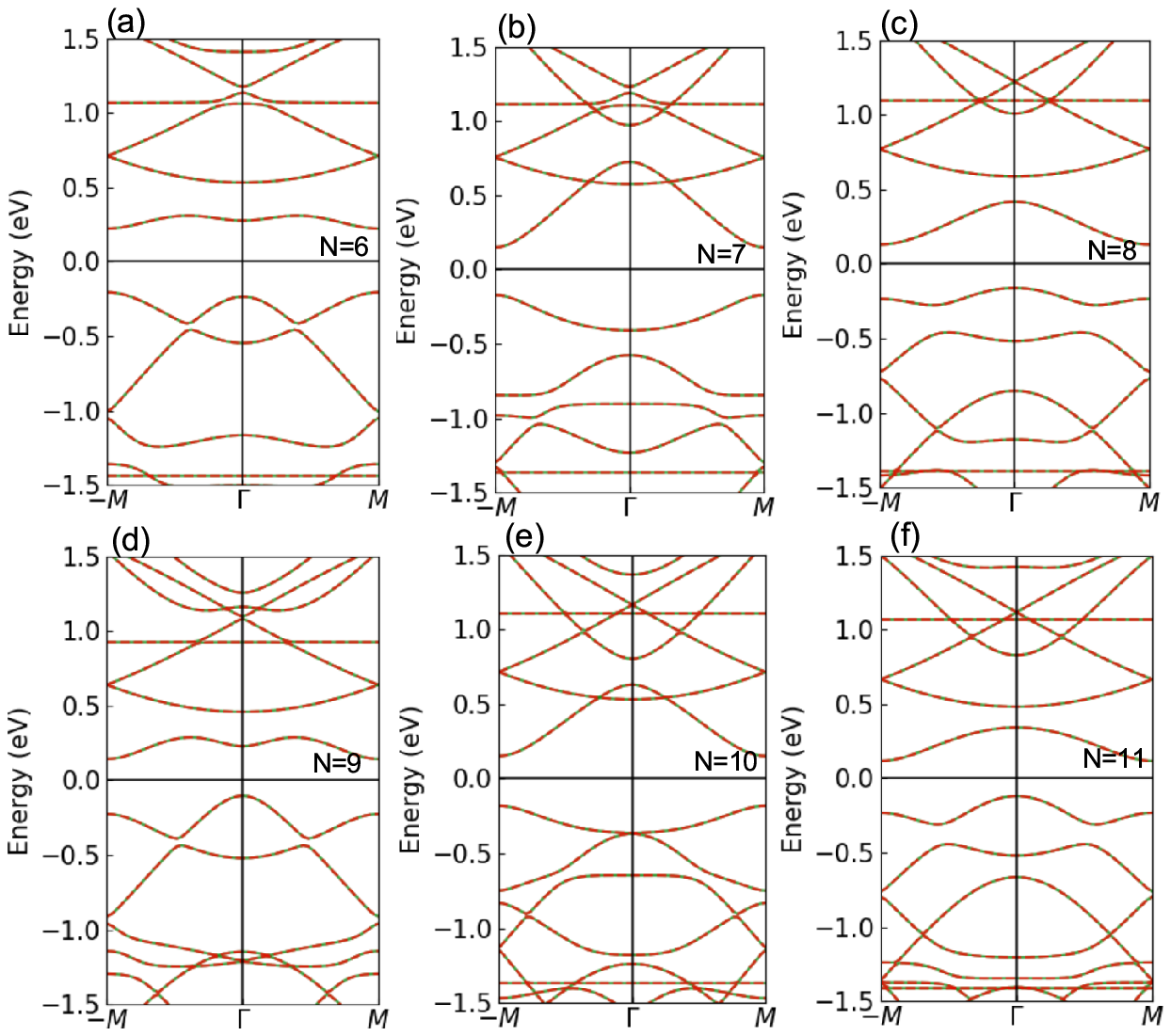}
\caption{\label{Fig4}
The spin-polarized band structures of $AH_{w}$ the ground-state phase for different ribbon widths:
$w$ = 6 (a), $w$= 7 (b), $w$=8 (c), $w$=9 (d), $w$=10 (e),  and $w$ = 11 (f).
The spin-up and spin-down bands are completely degenerate.}
\end{figure}

The spin-resolved band structures of the AFM state for the $AP_{w}$ with $w=$ 9, 10, 11, 12, 13 \ and \ 14 are shown in Figures 3(a-f), where one spin orientation is labeled as spin-up (green) and the opposite as spin-down (red). Obviously, the bands for all the cases are completely spin-degenerate. This is consistent with the spin density distribution ($\rho^{s}(r) = \Sigma_{n,k}(\mid \phi ^{\uparrow}_{nk}\mid^{2}-\mid \phi ^{\downarrow}_{nk}\mid^{2})$) in the AFM stable state, as shown for the $AP_{11}$ in Figure 1(a), where the $\mid \phi ^{\uparrow}_{nk}\mid^{2}$ and $\mid \phi ^{\downarrow}_{nk}\mid^{2}$ denote the electron density of spin-down (red) and spin-up (green), respectively. The main part of $\rho^{s}(r)$ is at the carbon atoms along the the pentagon edge. Moreover, the $AP_{w}$ nanoribbons have an indirect band gap across the Fermi level, with the conduction band minimum (CBM) located at the $M$ point and the valence band maximum (VBM) located at the $\Gamma$ point. This is different from the direct band gap at the $\Gamma$ point for pristine ANRs, meaning that the electronic property of pristine-ANRs can be tuned through edge reconstruction.

We present spin-resolved band structures of the $AH_{w}$ nanoribbons in the AFM  semiconductor states in Fig. 4, with $w=6$, 7, 8, 9, 10, and 11. In contrast to the $AP_w$ series, the $AH_{w}$ cases for $w = 8$, 9, and 11 are indirect band gap semiconductors with VBM at the $\Gamma$ point and CBM at the $M$ point, while the $AH_{w}$ cases for $w = 6$, 7, and  10 are direct band gap semiconductors with both VBM and CBM at the $M$ point. This VBM transition between the $\Gamma$ and $M$ points in the $AH_{w}$ cases can explain why the periodic oscillation is lost in the $\Delta E$ and $E_g$ curves in Figures 2(d,e). This difference between $AP_w$ and $AH_{w}$ nanoribbons can be attributed to their different edge structures: pure pentagon edge for $AP_w$ and hybrid (pentagon and hexagon) edge for $AH_{w}$. It is interesting the CBM is at the $M$ point for all the cases and the VBM at either $\Gamma$ or $M$ point, depending on the type and the $w$ value  of nanoribbons.

\begin{figure}[ht]
\centering
\includegraphics[width=0.3\textwidth]{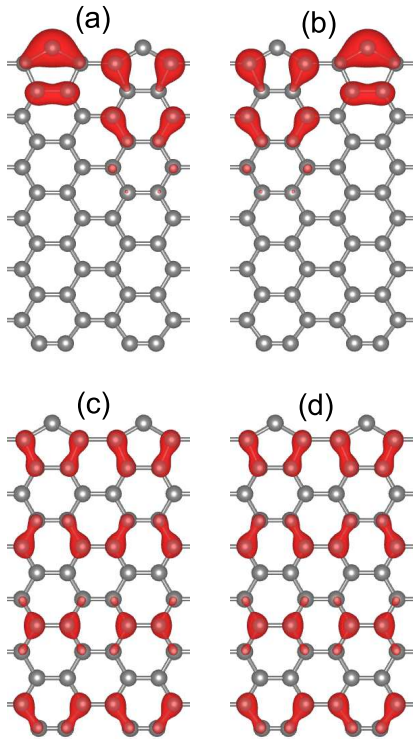}
\caption{\label{Fig5}
The real-space weight distributions of the degenerate band edges of the $AP_{12}$ in the ground-state phase. The conduction band edges at the $M$ point (a, b) indicate the spin-degenerate edge states along the pentagon edge, and the valence band edges at the $\Gamma$ (c, d) are spin-free bulk states similar to those of the pristine AGNRs. }
\end{figure}

It is interesting to further investigate the energy bands and determine their real-space weight distributions. For the $AP_{w}$ ground state, We plot the weight distributions of the band edges at the $\Gamma$ and $M$ points of the ground-state $AP_{12}$ in Figure 5, and present in Figure S2 those of the band edges (also valence maxima) of the $AP_{w}$ at the $\Gamma$ and $M$ points for the three widthes: $w=12$, 13, and  14. It is clear that the CBM states at the $M$ point are edge states along the pentagon edge and the highest valence states at the $M$ point also originate from the pentagon edge. In contrast, the VBM states at the $\Gamma$ point are from the whole bulk nanoribbons. For the $AH_{w}$ ground state, we plot the weight distributions in Figure S3 for $w=9$, 10, and 11. For $w=10$ (also 6 and 7), it is clear that the CBM and VBM states at the $M$ point originate from the hybrid edge. For $w=9$ and 11 (also 8), the energy bands near the $\Gamma$ and $M$ points and thus the band edge structures are similar to those of the $AP_{w}$ nanoribbons, and fortunately the real-space weight distributions are also similar to those of the $AP_{w}$ cases. Actually, the highest valence states at the $\Gamma$ point are the bulk states of the nanoribbons, and these  states at the $M$ point are the edge states originating from the pentagon or hybrid edge. For comparison, we also present real-space weight distributions of the band edges for the standard armchair nanorribons with the width $w=9$, 10, and 11 in Figure S4. It can be seen that the band edges are at the $\Gamma$ point and belong to the bulk states, except the conduction band edge for $w=3p+1$ (this band edge is higher than the CBM for both $w=3p$ and $w=3p+2$).

As a semiconductor, the $AP_{w}$ nanoribbon has semiconductor band structure with the CBM and VBM at the M and $\Gamma$ points in Brillouin zone. The spins are located at the edge carbon atoms. As for the $AH_{w}$, the CBM is also at the M point, but the VBM is located at the M  or $\Gamma$ points, depending the width $w$; and the spin is located at the outermost carbon atoms in the pentagons of the reconstructed edge. The spins make quantum Heisenberg antiferromagnetic spin-$\frac12$ chains described by Hamiltonian (1), where the exchange constant $J$ is equivalent to $\Delta E$.  In the thermodynamic limit, there is no long-range magnetic order in them because of the Mermin-Wagner theorem\cite{mermin}, but such quantum spin chains can host diffusive, super-diffusive, or ballistic spin transport, depending on the temperature and magnetization\cite{xxz1,xxz2,xxz3,xxz4}. More importantly, if the chain length becomes finite, such as $l=30$, or the magnetization is nonzero, there are nonzero spin Drude weights at finite temperature, and thus ballistic quantum spin transport can be achieved along the reconstructed edges\cite{xxz1,xxz2,xxz3,xxz4,spin1,spin2,spin3}. These ballistic spin transport could be used for quantum spin information transfer and spintronic applications.

\section{Conclusion}

In summary, we have achieved Heisenberg antiferromagnetic spin-$\frac12$ chains through edge reconstruction in AGNRs. The reconstructed edge consists of pentagonal carbon rings or a hybrid of pentagonal and hexagonal carbon rings. The resultant nanoribbons, with one edge reconstructed and the other remaining pristine armchair edge, are narrow-gap semiconductors and the band edge states are either spin-degenerate edge states or nonmagnetic bulk states. The spin originates from the $p_z$ electron of the outermost carbon of the pentagonal ring, and the inter-spin exchange is the nearest-neighbor antiferromagnetic interaction. For finite chain lengthes or nonzero magnetization, there are nonzero spin Drude weights in such one-dimensional quantum spin-$\frac12$ models, and thus ballistic quantum spin transport can be achieved along the reconstructed edges, These could be used for quantum spin information transfer and spintronic applications.


\begin{acknowledgments}
This work is supported by the Nature Science Foundation of China (Grant Nos.11974393 and 11574366) and the Strategic Priority Research Program of the Chinese Academy of Sciences (Grant No. XDB33020100). All the numerical calculations were performed in the Milky Way \#2 Supercomputer system at the National Supercomputer Center of Guangzhou, Guangzhou, China.
\end{acknowledgments}

%

\end{document}